\documentclass[conference]{IEEEtran}
\IEEEoverridecommandlockouts
\usepackage{acronym}
\usepackage{amsfonts}
\usepackage[dvips]{graphicx}
\usepackage{times}
\usepackage{cite}
\usepackage{amsmath}
\usepackage{array}
\usepackage{amssymb}
\usepackage{stfloats}
\usepackage{diagbox}
\usepackage{graphicx}
\usepackage{footnote}
\usepackage{amsthm}
\usepackage{booktabs}
\usepackage{array}
\usepackage[ruled,vlined]{algorithm2e}
\usepackage{subeqnarray}
\usepackage{cases}
\usepackage{threeparttable}
\usepackage{color}
\usepackage{epstopdf}
\usepackage{multirow}
\usepackage{tabularx}
\usepackage{enumerate}
\usepackage{multicol}
\usepackage{hyperref}
\usepackage{subfig}
\usepackage{caption}
\def\bb0{{\mathbb{0}}}

\def\ba{{\mathbf{a}}}
\def\bb{{\mathbf{b}}}
\def\bc{{\mathbf{c}}}

\def\bg{{\mathbf{g}}}
\def\bh{{\mathbf{h}}}

\def\bm{{\mathbf{m}}}
\def\bn{{\mathbf{n}}}

\def\bu{{\mathbf{u}}}

\def\bx{{\mathbf{x}}}
\def\by{{\mathbf{y}}}
\def\bz{{\mathbf{z}}}
\def\b0{{\mathbf{0}}}

\def\bH{{\mathbf{H}}}

\def\bU{{\mathbf{U}}}

\def\bW{{\mathbf{W}}}

\def\bbC{{\mathbb{C}}}

\def\cN{\mathcal{N}}

\def\cP{\mathcal{P}}

\def\sfD{\mathsf{D}}

\def\sfR{\mathsf{R}}

\def\sfU{\mathsf{U}}

\def\sfee{{\mathsf{e}}}

\def\sfp{{\mathsf{p}}}

\def\sfx{{\mathsf{x}}}

\def\sf0{{\mathsf{0}}}

\def\rm0{{\mathrm{0}}}

\def\j{\mathrm{j}}

\acrodef{CSI}[CSI]{channel state information}
\acrodef{CSIT}[CSIT]{channel state information at the transmitter}
\acrodef{CSIR}[CSIR]{channel state information at the receiver}
\acrodef{MIMO}[MIMO]{multiple-input multiple-output}
\acrodef{SISO}[SISO]{single-input single-output}
\acrodef{MISO}[MISO]{multiple-input single-output}
\acrodef{SIMO}[SIMO]{single-input multiple-output}
\acrodef{ADCs}[ADCs]{analog-to-digital convertors}
\acrodef{SNR}[SNR]{signal-to-noise ratio}
\acrodef{AWGN}[AWGN]{additive white Gaussian noise}
\acrodef{MRT}[MRT]{maximal ratio transmission}
\acrodef{DFT}[DFT]{Discrete Fourier Transform}
\acrodef{ULA}[ULA]{uniform linear array}
\acrodef{UPA}[UPA]{uniform planar array}
\acrodef{LS}[LS]{least squares}
\acrodef{ALMMSE}[ALMMSE]{approximate linear minimum mean squared error}
\acrodef{QIHT}[QIHT]{quantized iterative hard thresholding}
\acrodef{QIST}[QIST]{quantized iterative soft thresholding}
\acrodef{SVD}[SVD]{singular value decomposition}

\ifCLASSINFOpdf

\else
\fi
\begin{document}

\title{GNN-Enhanced Approximate Message Passing for Massive/Ultra-Massive MIMO Detection}
%\author{\IEEEauthorblockN{Hnegtao He}
%\author{\IEEEauthorblockN{Michael Shell}
%\IEEEauthorblockA{School of Electrical and\\Computer Engineering\\
%Georgia Institute of Technology\\
%Atlanta, Georgia 30332--0250\\
%Email: http://www.michaelshell.org/contact.html}
%\and
%\IEEEauthorblockN{Homer Simpson}
%\IEEEauthorblockA{Twentieth Century Fox\\
%Springfield, USA\\
%Email: homer@thesimpsons.com}
%\and
%\IEEEauthorblockN{James Kirk\\ and Montgomery Scott}
%\IEEEauthorblockA{Starfleet Academy\\
%San Francisco, California 96678--2391\\
%Telephone: (800) 555--1212\\
%Fax: (888) 555--1212}}
% make the title area

\author{

\IEEEauthorblockN{Hengtao He\IEEEauthorrefmark{1}, Alva Kosasih\IEEEauthorrefmark{2}, Xianghao Yu\IEEEauthorrefmark{3}, Jun Zhang\IEEEauthorrefmark{1}, \\ S.H. Song\IEEEauthorrefmark{1}, Wibowo Hardjawana\IEEEauthorrefmark{2}, and Khaled B. Letaief\IEEEauthorrefmark{1}}

\IEEEauthorblockA{\IEEEauthorrefmark{1}Department of ECE, The
Hong Kong University of  Science and  Technology, Kowloon, Hong Kong \\
E-mail: \{eehthe, eejzhang, eeshsong, eekhaled\}@ust.hk}

\IEEEauthorblockA{\IEEEauthorrefmark{2}Centre of Excellence in Telecommunications,
The University of Sydney, Sydney, Australia \\
E-mail: \{alva.kosasih, wibowo.hardjawana\}@sydney.edu.au}

\IEEEauthorblockA{\IEEEauthorrefmark{3}Department of Electrical Engineering, City University of Hong Hong,
E-mail: alex.yu@cityu.edu.hk}

%\IEEEauthorblockA{\IEEEauthorrefmark{3}Department of EIE, The Hong Kong Polytechnic University, Hong Kong,
%E-mail: jun-eie.zhang@polyu.edu.hk}
%\thanks{This work is supported by the Hong Kong Research Grant Council under Grant No. 16212120.}

\thanks{This work was supported by the General Research Fund (Projects No. 16212120, 16212922) and Research Impact Fund (Project No. R5009-21) from the Hong Kong Research Grants Council. The work was also supported by the Shenzhen Science and Technology Innovation Committee under Grant SGDX20210823103201006.
}

}
%
%\thanks{The work of Hengtao He and S. Jin was supported in part by the National Science Foundation (NSFC) for Distinguished Young Scholars of China with Grant 61625106 and the National Natural Science Foundation of China under Grant 61531011. The work of C.-K. Wen was supported by the ITRI in Hsinchu, Taiwan, and the MOST of Taiwan under Grants MOST 103-2221-E-110-029-MY3.}

%\author{Hengtao He, Chao-Kai Wen, Shi Jin,~and Geoffrey Ye Li
%
%\thanks{H.~He and S.~Jin are with the National Mobile Communications Research
%Laboratory, Southeast University, Nanjing 210096, China (e-mail: hehengtao@seu.edu.cn, and jinshi@seu.edu.cn).}
%\thanks{C.-K.~Wen is with the Institute of Communications Engineering, National
%Sun Yat-sen University, Kaohsiung 804, Taiwan (e-mail: chaokai.wen@mail.nsysu.edu.tw).}
%\thanks{G.~Y.~Li is with the School of Electrical and Computer Engineering,
%Georgia Institute of Technology, Atlanta, GA 30332 USA (e-mail:
%liye@ece.gatech.edu).}
%}

\maketitle

% As a general rule, do not put math, special symbols or citations
% in the abstract
\begin{abstract}
Efficient massive/ultra-massive multiple-input multiple-output (MIMO) detection algorithms with satisfactory performance and low complexity are critical to meet the high throughput and ultra-low latency requirements in 5G and beyond communications, given the extremely large number of antennas. In this paper, we propose a low complexity graph neural network (GNN) enhanced approximate message passing (AMP) algorithm, AMP-GNN, for massive/ultra-massive MIMO detection. The structure of the neural network is customized by unfolding the AMP algorithm and introducing the GNN module  for multiuser interference cancellation. Numerical results will show that the proposed AMP-GNN significantly improves the performance of the AMP detector and achieves comparable performance as the state-of-the-art deep learning-based MIMO detectors but with reduced computational complexity.  Furthermore, it presents strong robustness to the change of the number of users.
%In this paper, we consider Multiple-Input Multiple-Output detection by using deep learning approach. We propose a model-driven deep neural network, which is specially designed by unfolding the iterative algorithm. Some  trainable parameters are optimized through deep learning techniques aim to improve detection performance. Furthermore, this network are easy to train with few adjustable parameters, and can handle time-varying channel in single training. Numerical results show that the proposed method can improve the iterative algorithm efficiently
\end{abstract}

%\begin{IEEEkeywords}
%Deep learning, model-driven, MIMO detection, iterative algorithms, neural networks, graph neural networks
%\end{IEEEkeywords}

% no keywords
% For peer review papers, you can put extra information on the cover
% page as needed:
% \ifCLASSOPTIONpeerreview
% \begin{center} \bfseries EDICS Category: 3-BBND \end{center}
% \fi
%
% For peerreview papers, this IEEEtran command inserts a page break and
% creates the second title. It will be ignored for other modes.
\IEEEpeerreviewmaketitle

\section{Introduction}
Massive/Ultra-massive multiple-input multiple-output (MIMO) communication has become one of the enabling technologies for 5G and beyond networks owing to the high spectral efficiency and link reliability. Given the increasing size of antenna arrays, efficient MIMO detection algorithms are of significant importance to unleash the full potential of  MIMO systems, and a series of research has been conducted to balance the performance and complexity \cite{SD,AMPJSTSP,EPdetector}. To this end, iterative detectors based on approximate message passing (AMP) \cite{dynamicAMP} and expectation propagation (EP) \cite{EP} have been proposed. The AMP-based detector \cite{AMPJSTSP} is of low complexity and easy to implement in practice because only the matrix-vector multiplication is involved. However, it only favors the scenario when the elements of the MIMO channel matrix follow independent identically distributed (i.i.d.) sub-Gaussian distribution with zero mean. In contrast, the EP-based detector \cite{EPdetector} achieves Bayes-optimal performance when the channel matrix is unitarily invariant, but has a higher complexity than the AMP-based detector owing to involves a matrix inversion.
%However, for practical MIMO systems, the performance of these iterative detectors is still far from Bayes-optimal solution and has serious deterioration with correlated MIMO channels and imperfect channel state information (CSI).

Owing to its strong ability to extract useful features from data, deep learning (DL) has been recently utilized in the physical layer design of wireless communications \cite{Khlaed_JSAC,6G_khaled,Modeldriven18DL}, such as millimeter-wave channel estimation \cite{DL2018HE}, channel state information (CSI) feedback \cite{DL2018Wen}, and data detection \cite{MIMO_Detection20DL}. For MIMO detection, it has been shown that DL can improve traditional message passing detectors, such as AMP, and EP, with different strategies. Specifically, the OAMP-Net and OAMP-Net2 detectors \cite{MIMO_Detection20DL} were developed by unfolding the OAMP detector and introducing several learnable parameters. However, the performance improvements are still limited owing to small number of learnable variables.
%Such detectors were shown to achieve a significant performance improvement compared with the OAMP detector because DL can learn the optimal parameters  from the data. Furthermore, an RE-MIMO detector \cite{REMIMO20DL} that is based on the recurrent inference machines (RIM) \cite{RIM} was developed by integrating a self-attention mechanism into the iterative detector to reduce the bit-error rate.

Recently, graph neural networks (GNNs) have been applied to wireless communications \cite{YifeiGNN2}, such as to learn a message-passing solution for  inference problems \cite{Scotti20GNN,Kosasih20DL}. In particular, a GNN-based MIMO detector was developed by utilizing a pair-wise Markov random field (MRF) model \cite{Scotti20GNN}. As GNNs can capture the structure of the data into the feature vectors for the nodes and update them through message-passing methods, they are very promising solutions for enhancing  the performance of the message-passing-based detectors. For example, the GEPNet was developed by incorporating the GNN into the EP detector\cite{Kosasih20DL}. %The idea behind the GEPNet is to use GNN to increase the accuracy of the approximation in EP detector.
However, the computational complexity of GEPNet is still high because of the matrix inversion in each layer. An efficient DL-based MIMO detection algorithm, which strikes a good balance between the performance and complexity, is by far not available.

%Therefore, it is important to develop a low-complexity MIMO detector by using the GNN to enhance existing message passing detectors.

In this paper, we develop a model-driven DL-based MIMO detector, AMP-GNN, where the neural network structure is obtained by unfolding the AMP detector and incorporating the GNN module. In particular, the GNN module receives the equivalent AWGN observation from the AMP as input and outputs a refined version back to AMP at each layer. It can improve the accuracy of the equivalent AWGN observation for AMP algorithm. As a result,  it is helpful for multi-user interference (MUI) cancellation. The AMP-GNN also inherits the low-complexity of the AMP detector and improves the performance with the GNN module. Simulation results shall demonstrate that the proposed AMP-GNN significantly outperforms existing the AMP detector and achieve comparable performance as the state-of-the-art GEPNet detector while reducing the computational complexity.

\emph{Notations}---For any matrix $\mathbf{A}$, $\mathbf{A}^{T}$, $\mathbf{A}^{*}$, and ${ \mathrm{tr}}(\mathbf{A})$ will denote the transpose, conjugate, and  trace of $\mathbf{A}$, respectively. In addition, $\mathbf{I}$ is the identity matrix, $\mathbf{0}$ is the zero matrix, and $\mathbf{1}_n$ is the $n$-dimensional all-ones vector. A proper complex Gaussian distribution with mean $\boldsymbol{\mu}$ and covariance $\boldsymbol{\Omega}$ can be described by the probability density function (pdf):
\begin{equation*}
  \mathcal{N}_{\mathbb{C}}(\mathbf{z};\boldsymbol{\mu},\boldsymbol{\Omega})=\frac{1}{\mathrm{det}(\pi \boldsymbol{\Omega})}
  e^{-(\mathbf{z}-\boldsymbol{\mu})^{H}\boldsymbol{\Omega}^{-1}(\mathbf{z}-\boldsymbol{\mu})}
\end{equation*}
%and for a real Gaussian distribution, we have the pdf as
%\begin{equation*}
%\mathcal{N}(x;\mu, \sigma^{2})\triangleq\frac{1}{\sqrt{2\pi\sigma^{2}}}e^{-\frac{(x-\mu)^{2}}{2\sigma^{2}}}.
%\end{equation*}
%The remaining part of this paper is organized as follows. Section \ref{Problem}
%formulates the MIMO detection problem and introduces the AMP algorithm. Next, the AMP-GNN and GNN are provided in Section \ref{OAMP-Net} as well as the complexity is analyzed. Then, numerical results are presented in Section \ref{Simulation}. Finally, Section \ref{con} concludes
%the paper.

\section{Problem Formulation and Algorithm Review}\label{Problem}
In this section, we first formulate the MIMO detection problem by adopting a Bayesian inference. To better understand the proposed AMP-GNN, we will review the AMP-based MIMO detector.
\subsection{MIMO Detection}
 We assume an uplink multi-user MIMO (MU-MIMO) systems where the base station (BS) equips $M$ antennas sevres $N$ single-antenna users. The symbol vector $\bx\in\mathbb{C}^{N\times1}$ is transmitted over a Rayleigh fading channel $\bH \in \mathbb{C}^{M \times N}$. Each element of $\bH$ and $\bx$ are drawn from an i.i.d. complex Gaussian distribution and the $Q$-QAM constellation, respectively. The received signal $\by\in\mathbb{C}^{M\times1}$ is given by
\begin{equation}\label{eq1}
  \by=\bH \bx+\bn,
\end{equation}
where $\bn\sim \mathcal{N}_{\mathbb{C}}(0,\sigma^{2}\mathbf{I}_{M})$ is the additive white Gaussian noise (AWGN).
%Although DL has been extended to complex domain, it is basically performed in the real-valued domain.
%To better implement our proposed AMP-GNN detector, we consider an equivalent real-valued representation for the MIMO detection problem which is obtained by considering the real $\mathfrak{R}(\cdot)$ and imaginary $\mathfrak{I}(\cdot)$ parts separately. Denote $\mathbf{x}=[\mathfrak{R}(\bar{\mathbf{x}})^{T}, \mathfrak{I}(\bar{\mathbf{x}})^{T}]^{T}$, $\mathbf{y}=[\mathfrak{R}(\bar{\mathbf{y}})^{T}, \mathfrak{I}(\bar{\mathbf{y}})^{T}]^{T}$, $\mathbf{n}=[\mathfrak{R}(\bar{\mathbf{n}})^{T}, \mathfrak{I}(\bar{\mathbf{n}})^{T}]^{T}$, and
%\begin{equation}\label{eq2}
%\mathbf{H}=\left[\begin{array}{c c c}
%\mathfrak{R}(\bar{\mathbf{H}}) && -\mathfrak{I}(\bar{\mathbf{H}}) \\
%\mathfrak{I}(\bar{\mathbf{H}}) && \mathfrak{R}(\bar{\mathbf{H}})
%\end{array}\right].
%\end{equation}
%Then, the signal model in (\ref{eq1}) can be rewritten in terms of real vectors and matrix form as follows,
%\begin{equation}\label{eq3}
%\mathbf{y}=\mathbf{H}\mathbf{x}+\mathbf{n}.
%\end{equation}
Based on the Bayes' theorem, the posterior probability $p(\mathbf{x}|\mathbf{y},\mathbf{H})$ can be factorized as
\begin{equation}\label{eq4}
    p(\mathbf{x}|\mathbf{y},\mathbf{H})=\frac{p(\mathbf{y}|\mathbf{x},\mathbf{H})p(\mathbf{x})}{p(\mathbf{y}|\bH)}
 = \frac{p(\mathbf{y}|\mathbf{x},\mathbf{H})p(\mathbf{x})}{\int p(\mathbf{y}|\mathbf{x},\mathbf{H})p(\mathbf{x})d\mathbf{x}}.
\end{equation}
Given the posterior probability $p(\mathbf{x}|\mathbf{y},\mathbf{H})$, the Bayesian MMSE estimate is obtained by
\begin{equation}\label{MMSE_estimate}
\hat{\mathbf{x}}=\int \mathbf{x} p(\mathbf{x}|\mathbf{y},\mathbf{H})d\mathbf{x}.
\end{equation}

However, the Bayesian MMSE estimator is computationally intractable because the MMSE estimation in (\ref{MMSE_estimate}) involves a high-dimensional integral. In \cite{dynamicAMP}, the AMP algorithm was proposed as an efficient alternative to recover signal $\mathbf{x}$. We will introduce the AMP-based detector in the following subsection.
\subsection{AMP-Based MIMO Detector}\label{sec:AMP}
\begin{algorithm}[!t]
\caption{AMP-based MIMO detector}
\label{Algorithm:BAMP}
{
\begingroup
%\addtolength{\jot}{0.3em}
\textbf{1. Input:} $\by$, $\bH$, $\sigma^2$, $\cP(\bx)$. \\
\textbf{2. Initialization:} $\hat{x}_n^{(1)}=0$, $\hat{v}_n^{(1)}=\frac{N}{M}$, $Z_m^{(0)}=y_m$.\\
\textbf{3. Output:} $\hat{\bx}^{(T)}$.\\
\textbf{4. Iteration:} \\
\For{$t=1,\cdots, T$}
{
 \setlength\abovedisplayskip{0pt}
 \setlength\belowdisplayskip{0pt}
 \begin{subequations}
 \begin{align}
 V_m^{(t)}&=\sum_{n=1}^N |h_{mn}|^2\hat{v}_n^{(t)} \label{Equ:AMP1}\\
 Z_m^{(t)}&=\sum_{n=1}^Nh_{mn}\hat{x}_n^{(t)}-\frac{V_m^{(t)}(y_m-Z_m^{(t-1)})}{\sigma^2+V_m^{(t-1)}} \label{Equ:AMP2}\\
 \Sigma_n^{(t)}&=\left(\sum_{m=1}^M\frac{|h_{mn}|^2}{\sigma^2+V_m^{(t)}}\right)^{-1} \label{Equ:AMP3}\\
 r_n^{(t)}&= \hat{x}_n^{(t)}+\Sigma_n^{(t)}\sum_{m=1}^M\frac{h_{mn}^{*}(y_m-Z_m^{(t)})}{\sigma^2+V_m^{(t)}} \label{Equ:AMP4}\\
 \hat{x}_{n}^{(t+1)}& =\mathbb{E}\{x_n|r_n^{(t)},\Sigma_n^{(t)}\}
 \label{Equ:AMP5}\\
 \hat{v}_{n}^{(t+1)}& =\text{Var}\{x_n|r_n^{(t)}, \Sigma_n^{(t)}\}
 \label{Equ:AMP6}
 \end{align}
 \end{subequations}
}
\endgroup
}
\end{algorithm}
%\begin{align*}
%  \hat{x}_{i}^{(t+1)}&=\mathbb{E}\{x_i|p(GNN)\} \\
%  \hat{v}_{i}^{(t+1)}&=\text{Var}\{x_i|p(GNN)\}
%\end{align*}
\begin{figure*}
  \centering
  \includegraphics[width=14cm]{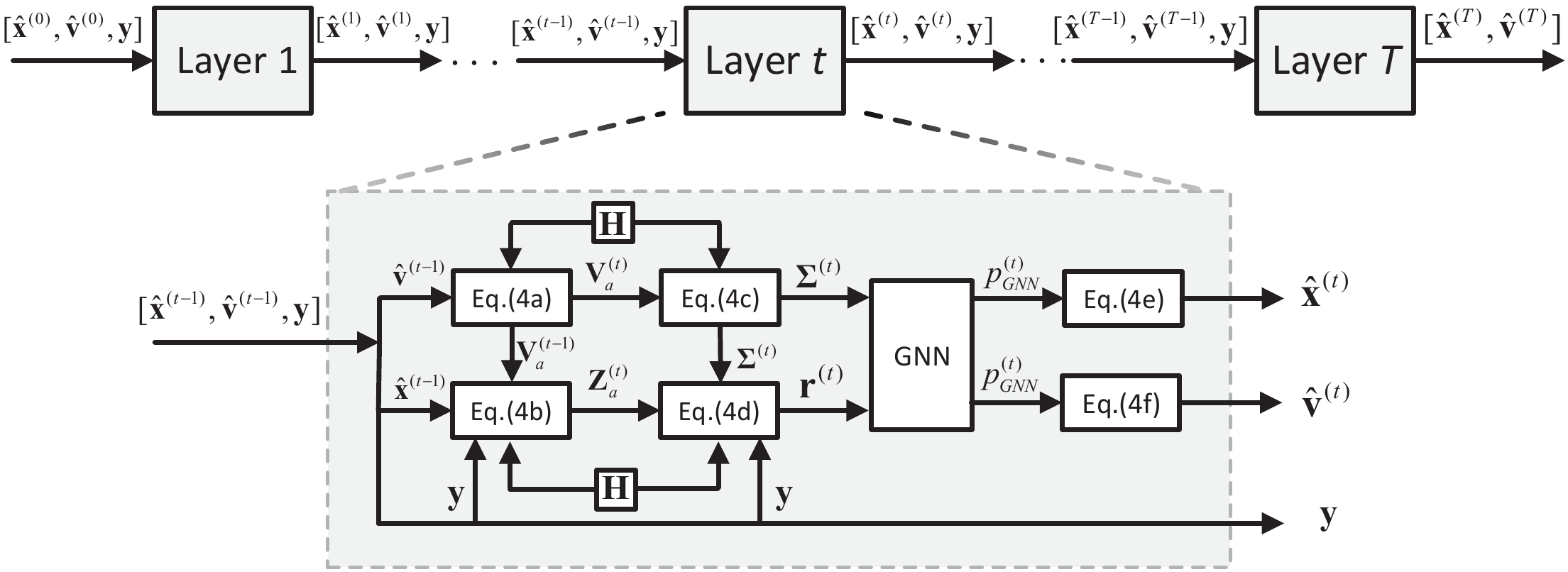}
  \caption{\small{.~~The structure of the proposed AMP-GNN detector.}}\label{fig:AMPGNN}
\end{figure*}

The AMP algorithm has been proposed to solve sparse linear inverse problems in compressed sensing \cite{dynamicAMP}. In Algorithm \,\ref{Algorithm:BAMP}, we summarize the AMP algorithm for MIMO detection\footnote{Note that we consider a complex-valued AMP-based MIMO detector in  Algorithm \,\ref{Algorithm:BAMP} and the equivalent real-valued form can be derived with the equivalent real-valued representation for (\ref{eq1}) accordingly.}, where $m$ and $n$ are the indexes of the antenna in the BS and users, respectively. The main principle of the algorithm is to decouple the posterior probability $p(\mathbf{x}|\mathbf{y},\mathbf{H})$ into a series of $p(x_{n}|\mathbf{y},\mathbf{H}), (n=1,2,\ldots,N)$, in an iterative way. In particular, $p(x_{n}|\mathbf{y},\mathbf{H})$ is assumed to be a Gaussian distribution and obtained from the equivalent AWGN model
\begin{equation}\label{eqAWGN}
    r_{n}^{(t)} = x_{n} + w_{n}^{(t)},
\end{equation}
% after computing  (\ref{Equ:AMP1}) and (\ref{Equ:AMP2}),
where  $w_{n}^{(t)}\sim \cN_{\mathbb{C}}(w_{n}^{(t)};0,\Sigma_n^{(t)})$.

To better understand Algorithm \ref{Algorithm:BAMP}, we first provide a brief explanation for each equation in Algorithm \ref{Algorithm:BAMP} with the detailed derivation given in \cite{dynamicAMP}. Specifically, (\ref{Equ:AMP1}) and (\ref{Equ:AMP2}) compute the variance and mean estimation for $\bz =\bH\bx$, respectively. Then,  (\ref{Equ:AMP3}) calculates the equivalent noise variance $\Sigma_n^{(t)}$ while
(\ref{Equ:AMP4}) gives the equivalent AWGN observation $r_n^{(t)}$. Finally, (\ref{Equ:AMP5}) and (\ref{Equ:AMP6}) perform the posterior mean and variance estimation for the equivalent AWGN model (\ref{eqAWGN}). As the transmitted symbol is assumed to be drawn from the $Q$-QAM set $\mathcal{S}=\{s_{1}, s_{2}, \ldots, s_{Q}\}$, the results in (\ref{Equ:AMP5}) and (\ref{Equ:AMP6}) are given by
\begin{equation}\label{eqmean}
  \hat{x}^{(t+1)}_{n} = \frac{\sum_{s_{i} \in \mathcal{S}}s_{i}\mathcal{N}_{\bbC}(s_{i};r_n^{(t)}, \Sigma_n^{(t)})p(s_{i})}{\sum_{s_{i} \in \mathcal{S}}\mathcal{N}_{\bbC}(s_{i};r_n^{(t)}, \Sigma_n^{(t)})p(s_{i})},
\end{equation}
\begin{equation}\label{eqvar}
  \hat{v}^{(t+1)}_{n} = \frac{\sum_{s_{i} \in \mathcal{S}}|s_{i}|^{2}\mathcal{N}_{\bbC}(s_{i};r_n^{(t)}, \Sigma_n^{(t)})p(s_{i})}{\sum_{s_{i} \in \mathcal{S}}\mathcal{N}_{\bbC}(s_{i};r_n^{(t)}, \Sigma_n^{(t)})p(s_{i})}
  -|\hat{x}^{(t+1)}_{n}|^{2}.
\end{equation}

As can be observed in Algorithm \,\ref{Algorithm:BAMP}, the performance of the AMP-based MIMO detector is mainly determined by the accuracy of the assumed equivalent AWGN model (\ref{eqAWGN}).  In \cite{dynamicAMP}, it was shown that the equivalent AWGN model is asymptotically accurate when the dimensions of the system tend to infinity, i.e.,  $M, N \rightarrow \infty$. However, in practical MIMO systems especially with high MUI, the performance of the AMP-based MIMO detector is far from optimal and even has an error floor owing to the inaccurate assumption. %Furthermore,  the existing state-of-the-art GEPNet detector requires matrix inversion, which is very complex.
These observations motivate us to improve the AMP-based MIMO detector with the DL technique, especially GNNs.
%\vspace{-0.5cm}
\section{AMP-GNN Detector}\label{OAMP-Net}
In this section, we propose an AMP-GNN detector for MIMO detection. First, we present the network structure of the AMP-GNN detector and introduce the GNN module in detail. Then, the computational complexity of the AMP-GNN is analyzed and compared with other MIMO detectors.
\subsection{AMP-GNN Architecture}
Given the important advantages of GNNs, such as learning to capture the MUI information, here we exploit them to improve the AMP detector in this article. The block diagram of the AMP-GNN is illustrated in Fig.\,\ref{fig:AMPGNN}. Specifically,  the network consists of $T$ cascade layers, and each layer has the same structure that contains a GNN module and a conventional AMP algorithm shown in Algorithm \,\ref{Algorithm:BAMP}. In particular, the input of the AMP-GNN is the received signal $\mathbf{y}$ and the initial value is set as $\hat{\mathbf{x}}^{(1)}=\mathbf{0}$ and $\hat{\mathbf{v}}^{(1)}=\frac{N}{M}\mathbf{1}_N$, and the output is the final estimate $\hat{\mathbf{x}}^{(T)}$ of signal $\mathbf{x}$. For the $t$-th layer of the AMP-GNN, the inputs are the estimated signal $\hat{\mathbf{x}}^{(t-1)}$ and $ \hat{\mathbf{v}}^{(t-1)}$ from the $(t-1)$-th layer and the received signal $\mathbf{y}$. %From Algorithm \,\ref{Algorithm:BAMP} and Fig.\,\ref{fig:AMPGNN}, the AMP-GNN mainly consists of an AMP algorithm and a GNN module.
As we have introduced the AMP-based MIMO in Section \ref{sec:AMP}, we will introduce  the GNN module in the next subsection.
\subsection{The GNN Module}
\begin{figure}[t]
  \centering
  \includegraphics[width=6cm]{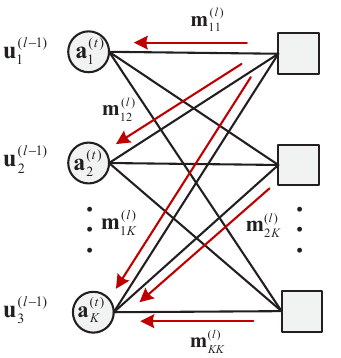}
  \caption{\small{.~~Message passing and update for GNNs.}}\label{fig:GNN}
\end{figure}

%As illustrated  in the Fig.\,\ref{fig:GNN}, there are two kinds of nodes in the GNN, i.e., the variable and factor nodes illustrated as circles and rectangles, respectively.
GNNs have been recently adopted in wireless communications as they can incorporate the graph topology of the wireless network into the neural network design \cite{YifeiGNN2}. The GNN model adopted in the AMP-GNN is called the message passing neural network (MPNN) and the message passing rules are illustrated in Fig.\,\ref{fig:GNN}. Similar to  \cite{Scotti20GNN}, the GNN in the AMP-GNN is composed of three main modules: a \emph{propagation module}, an \emph{aggregation module}, and a \emph{readout module}. The first two modules operate at all layers while the readout module is involved only after the last layer. To better understand the structure of the MPNN, we first elaborate the following concepts about the GNN.

\begin{itemize}
  \item \textbf{Node:} Each node $n \in V$ represents the $n$-th user.
  \item \textbf{Node attributes:} Each node has an assigned node attribute $\ba_{n}$ that is constant when exchange the information between nodes. In proposed AMP-GNN, the GNN in $t$-layer takes the output from the linear module in AMP algorithm as node attribute.
  \item \textbf{Edge:} An edge  $e_{n,j} \in E $ is to connect node $n \in V$  and $j \in V$. %Whether the edge exists or not depends on the graph structure.
  \item \textbf{Edge attributes:} Each edge $e_{n,j}$ has an assigned edge attribute $\mathbf{f}_{jn}$ that is constant when compute the message. In proposed AMP-GNN, the GNN use the CSI and noise level as edge attribute.
  \item \textbf{Hidden vector:} Each node $\bU_{n}$ has a hidden vector updated in the different round of the GNN, and will be used to compute the output of the GNN.
  \item \textbf{Message:} The incoming messages $\bm_{jn}$ from its connected edges are utilized to update the node feature vector.
\end{itemize}
%It can be adopted to model the correlation of the nodes in GNNs.

The reason for using GNN to model the correlation among the AWGN observations (\ref{eqAWGN}) is the connection between GNN and pair-wise MRF, which is investigated in the machine learning society  to model the structured dependency of a set of random variables $\bx = \{x_{1},\cdots,x_{N}\}$ by an undirected graph $G = \{V, E\}$. Specifically, the $n$-th variable node is characterized by a self potential $\phi(x_n)$, and the $(n,j)$-th pair of edge is characterized by a pair potential $\psi(x_n,x_j)$, which are given by
\begin{subequations}\label{phi_and_psi}
            \begin{equation}\label{phi}
\phi(x_n) = \sfee \sfx \sfp \left( \frac{1}{\sigma^2} \by^T \bh_n x_n - \frac{1}{2}  \bh_n^T \bh_n x_n^2  \right) p(x_n),
            \end{equation}
            \begin{equation}\label{psi}
\psi(x_n,x_j) = \sfee \sfx \sfp \left(- \frac{1}{\sigma^2} \bh_n^T \bh_j x_n x_j\right),
            \end{equation}
\end{subequations}
%Alva: Self and pair potentials must be defined close to when the term is first mentioned.
respectively, where $n,j \in \{1,\ldots, N\}$ and $n \neq j$. Finally, the joint probability $ p_{out}(\bx)$ corresponding to the pair-wise MRF can be obtained by GNN and written as \cite{Scotti20GNN}
\begin{equation}\label{p_x_y_GNN}
 p_{out}(\bx)(\bx) %{ \propto \mathcal{N} \left( \qy: \qH \qx ,  {\sigma}^2 \qI_{N_r} \right)  p(\qx) } \notag \\
 =\frac{1}{Z} \prod_{n=1}^N \phi(x_n) %{\underbrace{\mathcal{N} \left( x_k: x_{\rm obs,k}^{(t)} ,  v_{\rm obs,k}^{(t)} \right)}_{p(x_k)} }
 \prod_{\substack{j=1\\j \neq n}}^N \psi(x_n,x_j),
\end{equation}
where $Z$ is a normalization constant.

When designing the GNNs, we need to first define the node and edge attributes. As the GNN in the $t$-layer of the AMP-GNN takes  the output from the linear module in AMP algorithm as the input, it is natural to incorporate the mean $r_{n}^{(t)}$ and variance $\Sigma_n^{(t)}$  obtained from  (\ref{Equ:AMP3}) and (\ref{Equ:AMP4}) into the attribute $\ba_n^{(t)}$  of the variable node $x_n$ by concatenating the mean and variance as
\begin{equation}\label{concat}
\ba_n^{(t)} = \left[ r_{n}^{(t)}, \Sigma_n^{(t)} \right].
\end{equation}
% \begin{subequations}\label{phi_and_psi}
%             \begin{equation}\label{phi}
% \phi(x_k) ={ \sf exp} \left( \frac{1}{{\sigma}^2} \qy^T \qh_k x_k - \frac{1}{2}  \qh_k^T \qh_k x_k^2  \right) p(x_k),
%             \end{equation}
%             \begin{equation}\label{psi}
% \psi(x_k,x_j) = {\sf exp} \left(- \frac{1}{{\sigma}^2} \qh_k^T \qh_j x_k x_j\right).
%             \end{equation}
% \end {subequations}
To compute the joint probability $p_{out}(\bx)$ in \eqref{p_x_y_GNN}, we utilize the node and edge feature vectors corresponding to the self and pair potentials in \eqref{phi} and \eqref{psi}, respectively. The second step is to define the initialized hidden vector $\bu_n^{(\ell)}$ for each node $x_{n}$.  We consider the initial value is calculated from encoding the information of the received signal $\by$, corresponding channel vector $\bh_n$, and noise variance $\sigma^2$.  The encoding process is implemented by using a single layer neural network given by
\begin{equation}\label{GNN_init}
 \bu_n^{(0)} = \bW_1 \cdot [\by^T \bh_n, \bh_n^T \bh_n, \sigma^2]^T + \bb_1,
\end{equation}
where $\bW_1 \in \mathbb{R}^{N_u \times 3}$ is a learnable matrix, $\bb_1 \in \mathbb{R}^{N_u}$ is a learnable vector, and $N_u$ is the size of the feature vector.
The edge attribute  $\mathbf{f}_{jn} \triangleq \left[ \bh_n^T \bh_j, \sigma^2 \right] $ is obtained by extracting the pair potential information from \eqref{psi} and  utilized for the message passing of the GNN. %As illustrated in Fig.\,\ref{fig:GNN}, the initialized feature vectors are sent to the corresponding factor nodes. The factor nodes then commence the following operations in each module between the factor and variable nodes:
Next, we will elaborate the details of each module in the MPNN, including the propagation, the aggregation, and the readout module.
\subsubsection{\textbf{Propagation module}}
For any pair of variable nodes $x_n$ and $x_j$, there is an edge to connect them. Each edge first concatenates the connected hidden vectors $\bu_n^{(\ell-1)}$ and $\bu_j^{(\ell-1)}$ with its own attribute $\mathbf{f}_{jn}$ as
$\bc_{n}^{(l-1)} = [\bu_n^{{(}\ell-1{)}}, \bu_j^{{(}\ell-1{)}}, \mathbf{f}_{jn}]$. Then, it uses the concatenated features $\bc_{n}^{(l-1)}$ as the input for the multi-layer perceptron (MLP). Therefore, the output of the MLP is given by
\begin{equation}\label{factor_to_var}
\bm_{jn}^{(\ell)} = {\sfD} \left( \bc_{n}^{(l-1)} \right),
\end{equation}
where  $\sfD$ is the MLP network. In the propagation module, each edge has an MLP with two hidden layers of sizes $N_{h_1}$ and $N_{h_2}$ and an output layer of size $N_u$.  %In this paper, we set  $N_{h_1}=64$, $N_{h_2}=32$, and $N_u = 8$.
Furthermore, the rectifier linear unit (ReLU) activation function is used at the output of each hidden layer. Finally, the outputs $\bm_{jn}^{(\ell)}$ are fed back to the nodes as shown in Fig.\,\ref{fig:GNN}.
\subsubsection{\textbf{Aggregation  module}}
The $n$-th variable node  sums all the incoming messages $\bm_{jn}^{(\ell)}$ from its connected edges  and concatenates the sum of the $\bm_{jn}^{(\ell)}$  with the node attribute $\ba_n^{(t)}$ as $\bm_n^{(\ell)}  = \left[\sum_{\substack{j=1\\j \neq n}}^N \bm_{jn}^{(\ell)},  \ba_n^{(t)} \right]$. The $\bm_n^{(\ell)}$ is used to compute the node hidden vector $\bu_n^{(\ell)}$ as
 \begin{subequations}\label{GRU_MLP}
            \begin{equation}\label{GRU}
 \bg_n^{(\ell)} = {\sfU} \left( \bg_n^{(\ell-1)}, \bm_n^{(\ell)}  \right),
            \end{equation}
            \begin{equation}\label{MLP}
 \bu_n^{(\ell)}= \bW_2 \cdot \bg_n^{(\ell)} + \bb_2,
            \end{equation}
\end{subequations}
where the function $\sfU$ is specified by the gated recurrent unit (GRU) network, whose current and previous hidden states are $\bg_n^{(\ell)} \in \mathbb{R}^{N_{h_1} }$ and  $\bg_n^{({\ell}-1)} \in \mathbb{R}^{N_{h_1} }$, respectively. $\bW_2 \in \mathbb{R}^{N_u \times N_{h_1}}$ is a learnable matrix, and  $\bb_2\in \mathbb{R}^{N_u}$ is a learnable vector. The updated hidden vector \eqref{MLP} is then sent to the propagation module for next iterations.
\subsubsection{\textbf{Readout module}}
After $L$ rounds of the message passing between the propagation and aggregation module, a readout module is utilized to output the estimated distribution
$p_{\mathrm{GNN}}^{(t)}(x_n=s_{i}|\by)$ for the $t$-layer of AMP-GNN and given by
\begin{subequations}\label{Readout}
				\begin{equation}\label{Readout1}
\tilde p^{(t)}_{ \mathrm{GNN}}(x_n=s_{i}|\by)  = {\sfR} \left(  \bu_n^{(L)} \right), s_{i}\in \mathcal{S},
				\end{equation}
            	\begin{equation}\label{Readout2}
p_{\mathrm{GNN}}^{(t)}(x_n=s_{i}|\by)  = \frac{{\sfee \sfx \sfp } \left(\tilde p^{(t)}_{G\mathrm{NN}}(x_n=s_{i}|\by) \right)}  {\sum_{s_{i}\in \mathcal{S}}  {\sfee \sfx \sfp} \left({\tilde p}^{(t)}_{ \mathrm{GNN}}(x_n=s_{i}|\by) \right)}, s_{i}\in \mathcal{S}.
           		 \end{equation}
\end{subequations}
The readout function ${\sfR} $ consists of an MLP with two hidden layers of sizes $N_{h_1}$ and $N_{h_2}$, and ReLU activation is utilized at the output of each hidden layer. The output size of  ${\sfR} $ is the cardinality of the real-valued constellation set, i.e., $\sqrt{Q}$. Then, we compute
\begin{equation}\label{GNN_reset_index}
 \bg_n^{(0)}  \leftarrow   \bg_n^{(L)} \text { and }\bu_n^{(0)} \leftarrow   \bu_n^{(L)}, n= 1, \dots,N,
\end{equation}
for the next GNN iteration. Finally, we utilize $p_{\mathrm{GNN}}^{(t)}(x_n=s_{i}|\by)$ to compute the posterior mean and variance for the AMP, which are given by
\begin{subequations}\label{pyGNN}
            \begin{equation}\label{pyGNN1}
 \hat{x}_{n}^{(t+1)}=\mathbb{E}\{x_n|p_{\mathrm{GNN}}^{(t)}\},
            \end{equation}
            \begin{equation}\label{pyGNN2}
 \hat{v}_{n}^{(t+1)}=\text{Var}\{x_n|p_{\mathrm{GNN}}^{(t)}\}.
            \end{equation}
\end{subequations}
Note that the expectation and variance are computed with respect to $p_{\mathrm{GNN}}^{(t)}$. This is the main difference between AMP and AMP-GNN as $p_{\mathrm{GNN}}^{(t)}(x_n=s_{i}|\by)$ is not the Gaussian pdf in the AMP-GNN detector. After computing (\ref{pyGNN1}) and (\ref{pyGNN2}), the posterior mean $\hat{x}_{n}^{(t+1)}$ and $\hat{v}_{n}^{(t+1)}$ will be used for the next AMP-GNN iteration. Finally, the AMP-GNN is executed iteratively until terminated by a fixed number of layers.
\begin{table*}[t]	
\centering
%    \caption{\\LEARNABLE VARIABLES AND COMPUTATIONAL COMPLEXITY OF DIFFERENT DETECTORS.}
	\caption{Computational complexity (the number of multiplications) of different detectors. %The proposed AMP-GNN detector achieves a much lower complexity compared to the state-of-the-art DL-based MIMO detector GEPNet.
}
	\label{tab:complexity}
    \begin{tabular}{@{}lcccccc@{}}
    \toprule
    \diagbox{MU-MIMO settings}{Detectors}&OAMP-Net&GNN&GEPNet&AMP&AMP-GNN&RE-MIMO\\
    \midrule
    \midrule
%    Computational  complexity &$O(M{N}^2)$&$O(M{N}^2)$&$N_{h1}N_{h2}$&$O(M{N}^2)$&$O(MN)$& $O(MN)$ \\
%    \midrule
    $64 \times 64$ &$8.22\times10^{6}$&$1.17\times10^{6}$&$5.11\times10^{6}$&$1.78\times10^{5}$&$2.35\times10^{6}$& $4.93\times10^{8}$ \\
    \midrule
    $256 \times 256$ &$5.21\times10^{8}$&$1.27\times10^{7}$&$2.02\times10^{8}$&$2.68\times10^{6}$&$1.93\times10^{7}$& $3.22\times10^{9}$ \\
    \midrule
    $1024 \times 1024$ &$3.33\times10^{10}$&$5.56\times10^{8}$&$1.24\times10^{10}$&$4.22\times10^{7}$&$6.14\times10^{8}$& $4.41\times10^{10}$ \\
    \bottomrule
    \end{tabular}
\end{table*}

\subsection{Complexity Analysis}
We analyze the computational complexity of the AMP-GNN and compare it with existing massive MIMO detectors. Specifically,
the complexity of the AMP detector is  $\mathcal{O}(MN)$ owing to the matrix-vector multiplication while the complexity for GNN is  $\mathcal{O}(N_{h1}N_{h2})$ accounts for the each MLP operation. Therefore, the computational complexity of the AMP-GNN is $\mathcal{O}(MN+NN_{h1}N_{h2})$, dominated by the complexity of the AMP and GNN. In contrast, the complexity of the GEPNet is $\mathcal{O}(M{N}^2+NN_{h1}N_{h2})$ which includes the computational complexity of the EP and GNN.

To better compare the complexity, we use the number of multiplication as the metric and show the exact values for different MIMO settings with quadrature phase shift keying (QPSK) symbols in Table.\,\ref{tab:complexity}. The hyperparameters for GNN are set as $N_{h_1}=16$, $N_{h_2}=8$, and $N_u = 8$.  Compared with state-of-the-art DL-based MIMO detectors, such as  GEPNet and RE-MIMO, the AMP-GNN entails a much lower complexity. In particular, the ratio between the complexity of the AMP-GNN and GEPNet is dramatically reduced when the number of user increases. For example, the ratio between the complexity of the AMP-GNN and GEPNet is only $45.99\%$ when $M=N=64$  while the ratio is $4.95\%$ when $M=N=1024$.  This is because the complexity of matrix inversion  in the GEPNet will be the dominant term, which is prohibitively high when the number of antennas is large. In contrast, the AMP-GNN only involves  matrix-vector multiplications, which is a favorable feature for future ultra massive MIMO systems.
%Specifically,  the complexity of the AMP is $MN$ owing to the matrix multiplication while the complexity for GNN is  $N_{h1}N_{h2}$ owing to the MLP operation is required in GNN. Furthermore, the complexity of the OAMP, OAMP-Net and EP detectors in each layer is $MN^{2}$ because of matrix inversion is required. On the contrary, the complexity of GEPNet is $O(M{N}^2+N_{h1}N_{h2})$, including the computational complexity of the EP and GNN. When the number of antennas $M$ and the users $N$ is very large, the computational complexity of the AMP-GNN is much lower than  OAMP, OAMP-Net2, EP and GEPNet detectors, which is an advantageous feature for future massive MIMO system.
\section{Simulation Results}\label{Simulation}
In this section, we provide simulation  results of the AMP-GNN for massive MIMO detection and compare it with other MIMO detectors. We use the symbol error rate (SER) as the performance metric in our simulations. The signal-to-noise (SNR) of the system, defined as $\mathrm{SNR}=\frac{\mathbb{E}\|\mathbf{H}\mathbf{x}\|^{2}_{2}}{\mathbb{E}\|\mathbf{n}\|^{2}_{2}}$. To illustrate the effectiveness of our proposed AMP-GNN, we adopt several well established MIMO detectors as baselines:
\begin{itemize}
  \item \textbf{MMSE}: A classical linear receiver for MIMO detection which inverts the received signal by applying the channel-noise regularized pseudo-inverse
        of the channel matrix.
  \item \textbf{AMP}: An efficient message passing algorithm for MIMO detection given in Algorithm \ref{Algorithm:BAMP} and implemented with $10$ iterations\footnote{It was found that a  further increase in the number of iterations only offers a negligible performance gain. We set the same number of layers in other DL-based baseline methods for fair comparison.}.
 % \item \textbf{OAMP}: The OAMP-based MIMO detector with 10 iterations developed in \cite{MIMO_Detection20DL}.
  \item \textbf{OAMP-Net}: The OAMP-based model-driven DL detector developed in \cite{MIMO_Detection20DL}. Each layer requires computing a matrix pseudo-inverse and has 2 learnable parameters.
  \item \textbf{EP}: The EP-based MIMO detector with 10 iterations as proposed in \cite{EPdetector}.
  %\item \textbf{GNN}: The GNN-based MIMO detector  proposed in \cite{Scotti20GNN}.
  \item \textbf{GEPNet}: The GNN-enhanced EP detector proposed in \cite{Kosasih20DL} with $10$ layers.
 % \item \textbf{RE-MIMO}: The RE-MIMO detectors  proposed in \cite{REMIMO20DL} based on the RIM with $10$ layers.
 %  \item \textbf{ML}: The optimal detector implemented by Gurobi optimization package.
\end{itemize}
\begin{figure}
\begin{minipage}{3in}
  \centerline{\includegraphics[width=3.0in]{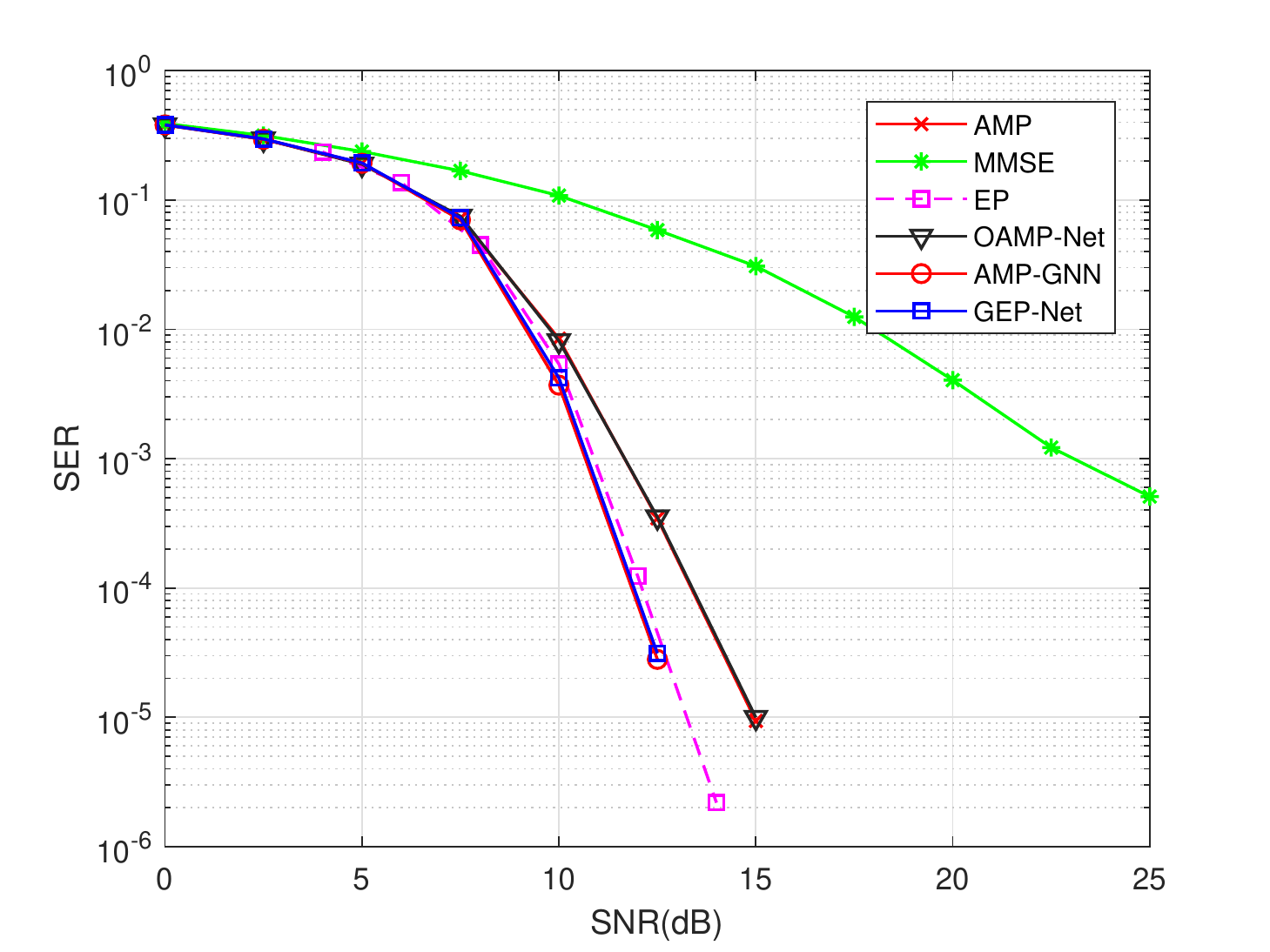}}
  \centerline{(a) QPSK}
\end{minipage}
\hfill
\begin{minipage}{3in}
  \centerline{\includegraphics[width=3.0in]{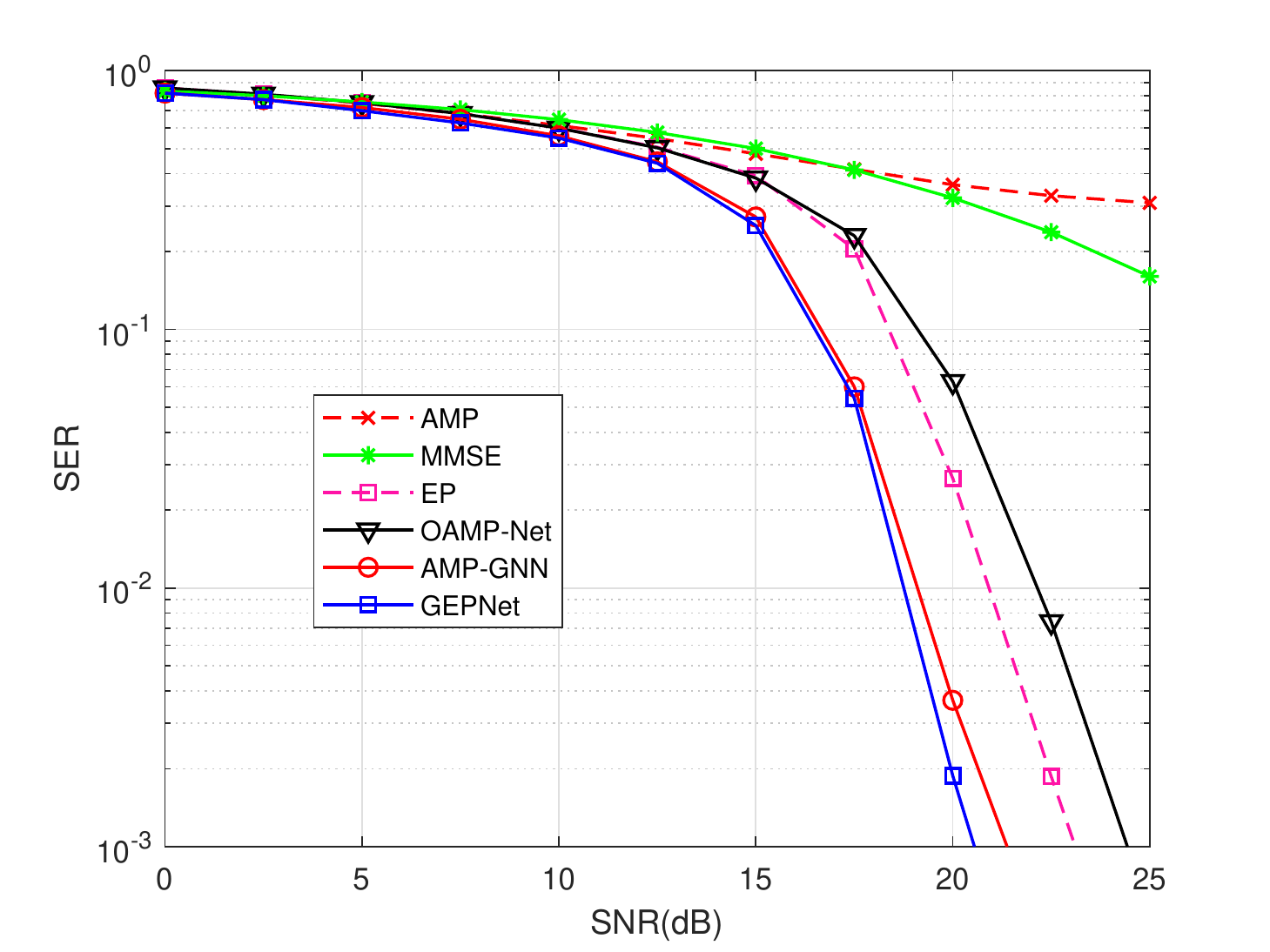}}
  \centerline{(b) 16-QAM}
\end{minipage}
\caption{.~~SER comparison of AMP-GNN with other MIMO detectors under $64\times64$ Rayleigh MIMO channels with
QPSK and 16-QAM symbols.}
\label{Fig:rayleigh}
\end{figure}
\subsection{Implementation Details}
In the simulation, the AMP-GNN is implemented on the PyTorch platform. The number of layers of the AMP-GNN detector  is set to $T=10$ while the number of layers of the GNN is set to $L=2$. The training data consists of a number of randomly generated pairs $(\mathbf{x},\mathbf{y})$. The data $\mathbf{x}$ is generated from QAM modulation symbols. We train the network with $100$ epochs. At each epoch, the training and validation sets contain $100,000$, and $5,000$ samples, respectively. The AMP-Net is trained using the stochastic gradient descent method and Adam optimizer. The learning rate is set to be $0.001$ and the batch size is set to $64$.
We choose the $L_{2}$ loss as the cost function, which is defined by,
\begin{equation}\label{eqloss}
  L_{2}(\mathbf{x},\hat{\mathbf{x}}^{(T)})=\|\mathbf{x}-\hat{\mathbf{x}}^{(T)}\|^{2}.
\end{equation}
%Furthermore, we consider the channel $\mathbf{H}$ is time-varying and each element drawn from $\mathbf{H}\sim\mathcal{N}_{\mathbb{C}}(0,1/M)$.
\subsection{Performance Comparsion}
Fig.\,\ref{Fig:rayleigh} compares the average SER of the AMP-GNN with those of the baseline detectors under $64\times64$ Rayleigh MIMO channels with
QPSK and $16$-QAM symbols, respectively. As can be observed from the figures, the AMP-GNN outperforms almost all other MIMO detectors except for the GEPNet detector. In particular, the AMP-GNN outperforms the AMP detector in all SNRs, which demonstrates the GNN can enhance traditional iterative detector with the help of information from data. Compared with  the state-of-the art GEPNet detector, the AMP-GNN can achieve similar performance but with lower complexity. Interestingly, the AMP-GNN can outperform EP detector but with lower complexity, which illustrates the GNN can help to avoid matrix inversion and provide a new thinking for developing message passing algorithms.
%Specifically, if we target an SER=$10^{-2}$, then the performance gain is approximately $3.9$ dB compared to the AMP detector. The reason for the performance improvement is that the GNN can refine the equivalent AWGN model with a more accurate distribution $p_{\mathrm{GNN}}^{(t)}(x_n=s_{i}|\by)$. Furthermore, the AMP-GNN has only $0.8$ dB performance loss compared to the GEPNet detector in $4\times4$ MIMO systems  when we target at an SER=$10^{-2}$. The performance loss then reduces to $0.3$ dB for $32\times32$ MIMO systems as illustrated in Fig.\,\ref{fig2_2}.%Thus, we can conclude the AMP-GNN has a comparable performance to GEPNet but with a remarkably reduced computational complexity, especially for massive and ultra-massive MIMO systems.
%Although the AMP-GNN has a slightly worse performance than the GEPNet,  it has a much lower complexity.
%For example, the AMP-GNN achieves a  $54.01\%$ complexity reduction compared with GEPNet in $64\times64$ MIMO systems as shown in Table.\,\ref{tab:complexity}.
%\begin{figure}
%  \centering
%  \includegraphics[width=3 in]{Fig_large}
%  \caption{\small{.~~SER comparison of AMP-GNN with other MIMO detectors under $32\times32$ Rayleigh MIMO
%   channels with QPSK symbols.}}\label{fig2_2}
%\end{figure}
\subsection{Robustness to Dynamic Numbers of Users}
\begin{figure}
  \centering
  \includegraphics[width=3in]{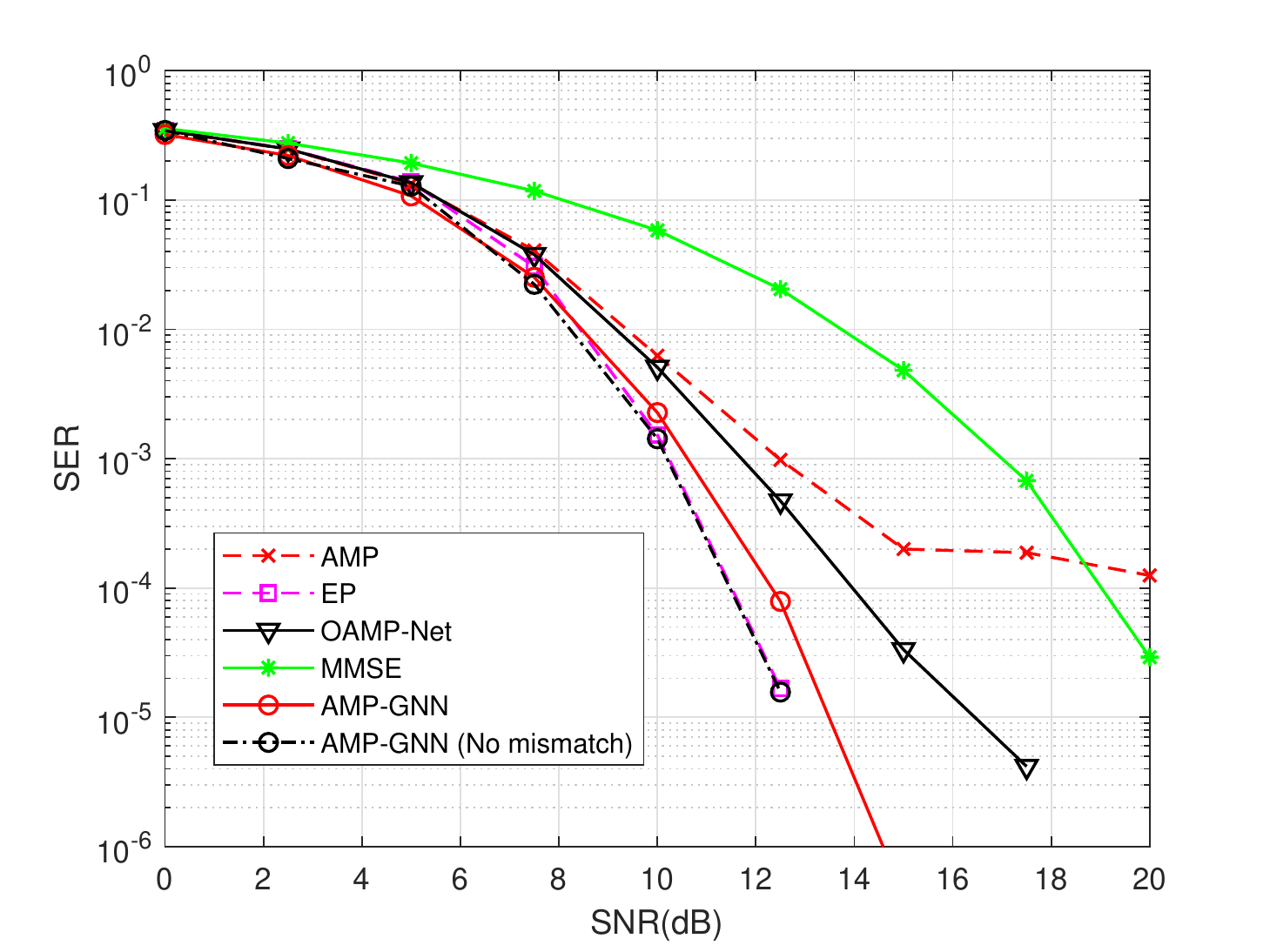}
  \caption{\small{.~~Robustness of the AMP-GNN for dynamic number of users.}}\label{fig3}
\end{figure}

Most of the DL-based MIMO detectors are trained and tested with a fixed number of antennas. However, the number of users in practical MIMO systems may quickly change. For example, new users may access to be served by the BS. As a result, the DL-based MIMO detector should have the ability to handle a varying number of users with a single model. In Fig.\,\ref{fig3}, we train the AMP-GNN in an $32\times16$  and $32\times32$ MIMO systems and test it in a $32\times24$ MIMO system. As shown in the figure, if we target an $\mathrm{SER}=10^{-3}$,  then the AMP-GNN still has $2.0$ dB performance gain compared with the conventional AMP detector even when tested with different number of users. Furthermore, it has a similar performance as the AMP-GNN trained and tested both in the $32\times24$ MIMO system. As a result, the AMP-GNN has strong robustness to different numbers of users in the deployment stage. This is because the GNN has the permutation equivariance property which makes it robust against the dynamic changes of the number of users.   %$N\in [N_{\mathrm{min}}, N_{\mathrm{max}}]$
%illustrates the SER performance of the AMP-GNN and other MIMO detectors with Rayleigh MIMO channels.
%In that case, all algorithms have performance degradation compared with the independent Rayleigh channel. For example, approximately $6.34$ dB and $6.05$ dB loss are caused due to the channel correlation for the OAMP-Net and OAMP algorithm, respectively, when we target the SNR for BER=$10^{-2}$. Furthermore,  the OAMP-Net can obtain more performance gain even if the channel has correlation. Compared with $1.86$ dB gain obtained in the independent Rayleigh channel, the OAMP-Net can obtain more than $2.15$ dB performance improvement under the correlated MIMO channel.
%Our OAMP-Net is superior to the TISTA, especially under the independent Rayleigh channel, as we introduce more trainable parameters in the network.
\section{Conclusions}\label{con}
We have developed a novel GNN-enhanced AMP detector for massive/ultra-massive MIMO detection, named AMP-GNN, which is obtained by incorporating the GNN module into the AMP algorithm. The AMP-GNN inherits the respective superiority of the low-complexity AMP detector and efficient GNN module. Simulation results demonstrated that  AMP-GNN improves the performance of the AMP detector significantly and has comparable performance as the state-of-the-art GEPNet detector but with a reduced computational complexity. Furthermore, it is quite robust to the change of the number of users in practice.
\section*{Acknowledgment}
The authors would like to thank Prof. Chao-Kai Wen, from the National
Sun Yat-sen University, for the discussion of neural enhanced message passing.%, and Prof. Shi Jin, from the Southeast University, for providing the GPU resources.

\end{document}